\documentclass[proceedings, preprint]{rmaa}

\usepackage{paralist}

\usepackage{psfrag,color}
\usepackage{comment}
\usepackage{xurl}

\usepackage{csquotes}

\usepackage{caption}
\SetYear{2022}
\SetConfTitle{Prospects for low-frequency radio astronomy in S. A.}

\title{Astrophysics as a Service: Turning radio astronomy as an opportunity to impact society and businesses} 

\author{Elias S. Fliger,\altaffilmark{1} Leandro M. García,\altaffilmark{1} and Martín Salibe\altaffilmark{1}}

\altaffiltext{1}{Instituto Argentino de Radioastronomía (CONICET -- CIC -- UNLP), C.C.5, (1984) Villa Elisa, Buenos Aires, Argentina (esfliger@iar.unlp.edu.ar).}

\shortauthor{Fliger, García, \& Salibe}
\shorttitle{Astrophysics as a Service}

\listofauthors{E. S. Fliger, L. M. García, \& M. Salibe}
\indexauthor{Fliger, E. S.}
\indexauthor{García, L. M.}
\indexauthor{Salibe, M.}

\abstract{For more than 25 years, the Instituto Argentino de Radioastronomía has been directing efforts from basic research and radio astronomy development to technology transfer projects around Argentina's National Space Plan and to Small and Medium Enterprises. With the surge of COVID-19, our organization’s transformation accelerated, bringing new opportunities and challenges which can be applied to impact health, education, processes and businesses. In this article, we explore our efforts to bridge the gap between basic science and the needs of our society.}

\resumen{Durante más de 25 años, el Instituto Argentino de Radioastronomía ha dirigido sus esfuerzos hacia el desarrollo de proyectos de transferencia de tecnología, en estrecha relación con el Plan Espacial Nacional de la Comisión Nacional de Actividades Espaciales y el desarrollo productivo de las pequeñas y medianas empresas. Con el advenimiento de la pandemia de COVID-19, se ha acelerado el proceso de transformación organizacional, generando nuevas oportunidades y desafíos que pueden aplicarse en áreas como la salud, la educación y los procesos empresariales. En este artículo, exploramos los esfuerzos realizados para cerrar la brecha entre la ciencia básica y las necesidades de nuestra sociedad. }

\addkeyword{Technology transfer}
\addkeyword{Open innovation}
\addkeyword{Radio astronomy}
\addkeyword{Social impact}
\addkeyword{Businesses}

\begin{document}
% Typeset article header
\maketitle

\section{Introduction}
\label{sec:intro}
The Instituto Argentino de Radioastronomía (IAR) is a renowned research institution founded in 1962 in Argentina. It was established through a collaboration between the Consejo Nacional de Investigaciones Científicas y Técnicas (National Council for Scientific and Technical Research, CONICET), the Comisión de Investigaciones Científicas of the Buenos Aires province (Scientific Research Commission, CIC), the Universidad Nacional de La Plata (UNLP), and the Universidad de Buenos Aires (UBA). The primary objective of the IAR is to advance research and technical development in the field of radio astronomy in Argentina and South America, while also contributing to astrophysics, education, and public outreach.
    
\begin{figure}[!t]
  \includegraphics[width=\columnwidth]{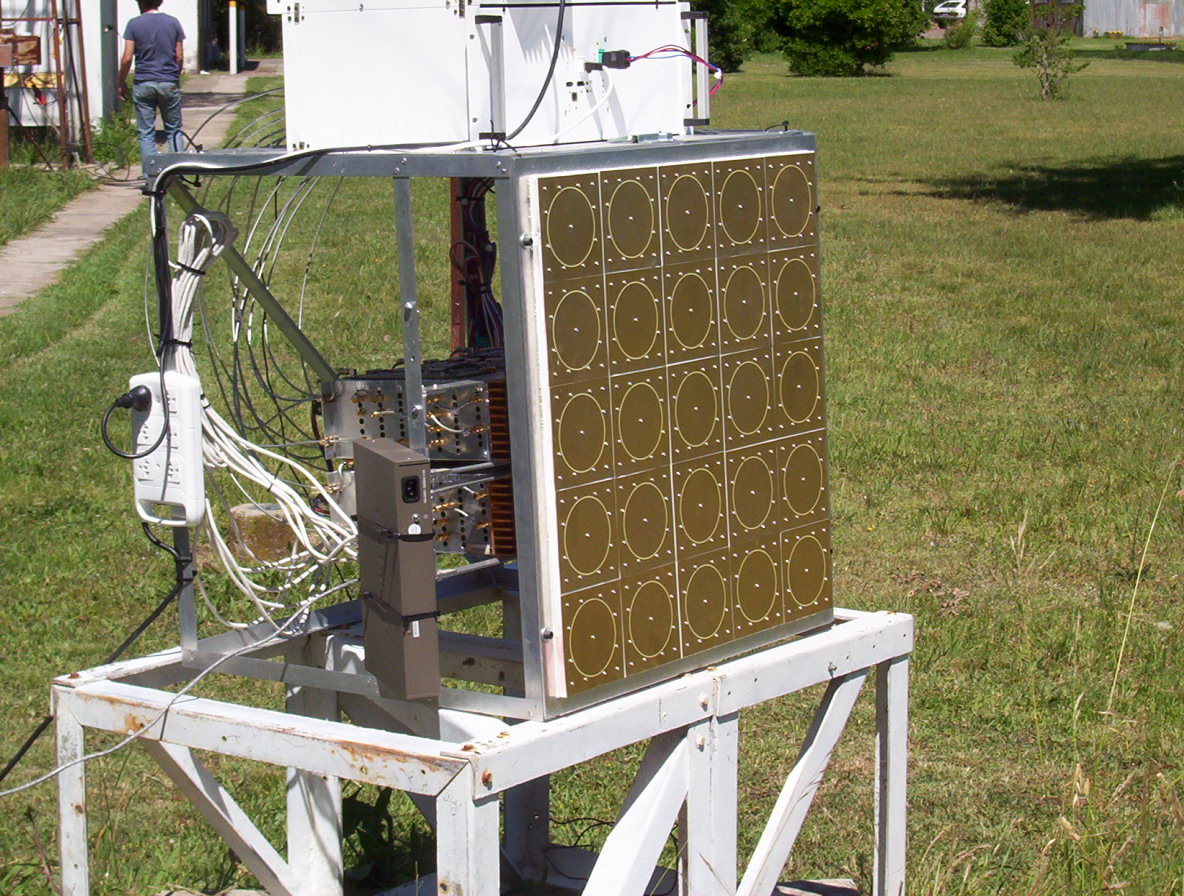}
  \caption{SAOCOM SAR antenna prototype in 2003.}
  \label{fig:saocom-sar}
\end{figure}

In the late 90s, IAR became a key player in Argentina's National Space Plan and began its technology transfer (TT) initiatives, forging collaborations with the Comisión Nacional de Actividades Espaciales (CONAE), the country's space agency. IAR's first technology transfer initiatives involved the development of the first Synthetic Aperture Radar (SAR) prototypes for the Argentinian L-Band satellite mission SAOCOM-1 (see Fig. \ref{fig:saocom-sar}), an airborne system that was developed as a first conceptual step toward the mission's design \citep{Skora2005}.

\begin{figure}[!t]
  \includegraphics[width=\columnwidth]{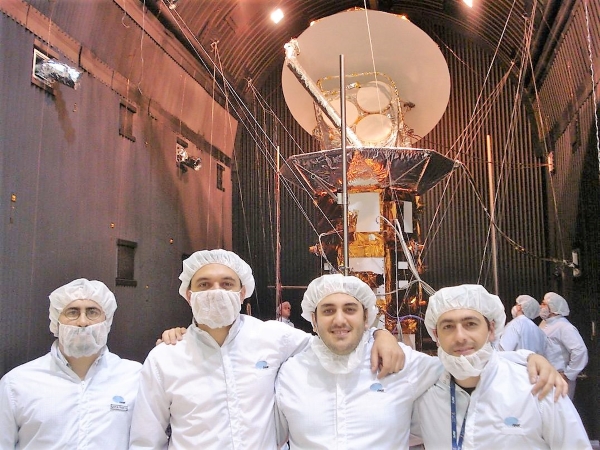}
  \caption{IAR's engineering team in front of the SAC-D/Aquarius satellite before Thermal-Vaccum tests.}
  \label{fig:iar_sacd}
\end{figure}

\begin{figure}[!t]
  \centering
  \includegraphics[width=0.75\columnwidth]{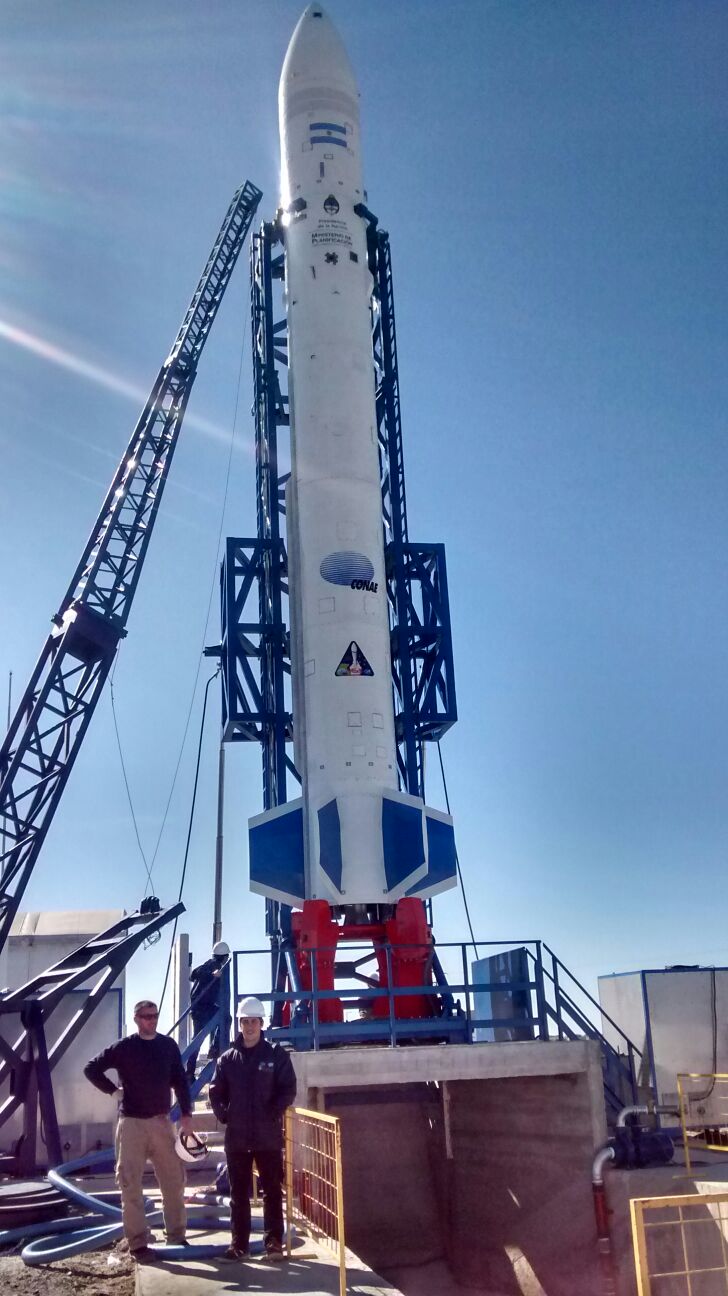}
  \caption{Vex 1A experimental launcher and IAR's technical staff, Pablo Alarcón and Eliseo Díaz, during the tests campaign.}
  \label{fig:vex1a}
\end{figure}

\begin{figure}[!t]
  \includegraphics[width=\columnwidth]{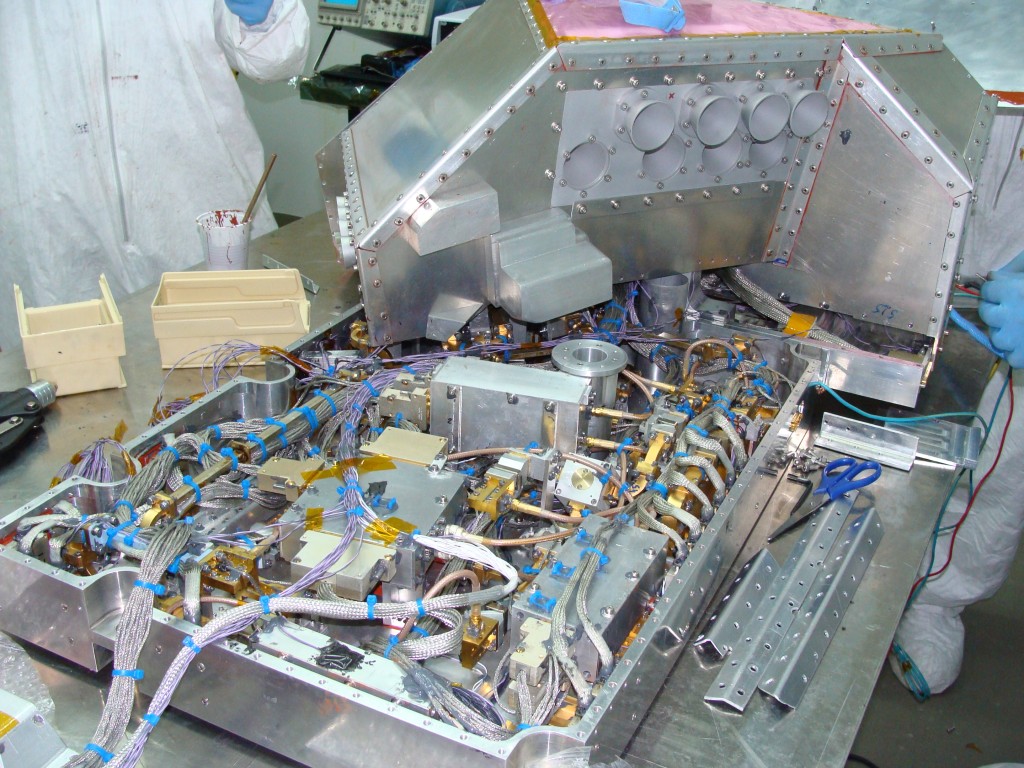}
  \caption{MWR during its assembly and integration to the SAC-D/Aquarius satellite.}
  \label{fig:mwr}
\end{figure}

These initiatives paved the way for IAR's involvement in significant projects, such as the SAC-D/Aquarius satellite mission (see Fig. \ref{fig:iar_sacd}), by contributing to the development of two main instruments: the MicroWave Radiometer (MWR) (see Fig. \ref{fig:mwr}) and the New InfraRed Sensor Technology (NIRST), the design and development of several subsystems such as the Data Acquisition Platform (PAD), and antennas aimed for communication and GPS. Additionally, IAR contributed to other aerospace projects like the experimental series of launchers Vex for the Tronador II program (see Fig. \ref{fig:vex1a}), and the electronic design for the Thermal InfraRed (TIR) camera of the SABIAMar mission.

Beyond space-related endeavors, IAR's TT activities have extended to various industry sectors. For instance, the institute is making notable contributions to the medical field, including the development of the Image based Microwave Tomography (ITM) still in its prototype stage. During the COVID-19 pandemic, researchers, engineers, and technicians, swiftly assembled a rapid response team that successfully designed an ozone reactor together with a local Small and Medium-sized Enterprise (SME), and proposed a turbine based Non-Invasive Mechanical Ventilator (NIMV). More recently, new government funded subsidies, promoting associations between the science and technology system and worker cooperatives, lead to the design and development of a Low-Cost Industrial Internet of Things (IIoT) network of devices for a 120 Tn/day vegetable oil refinery.

\begin{figure}[!t]
    \includegraphics[width=\columnwidth]{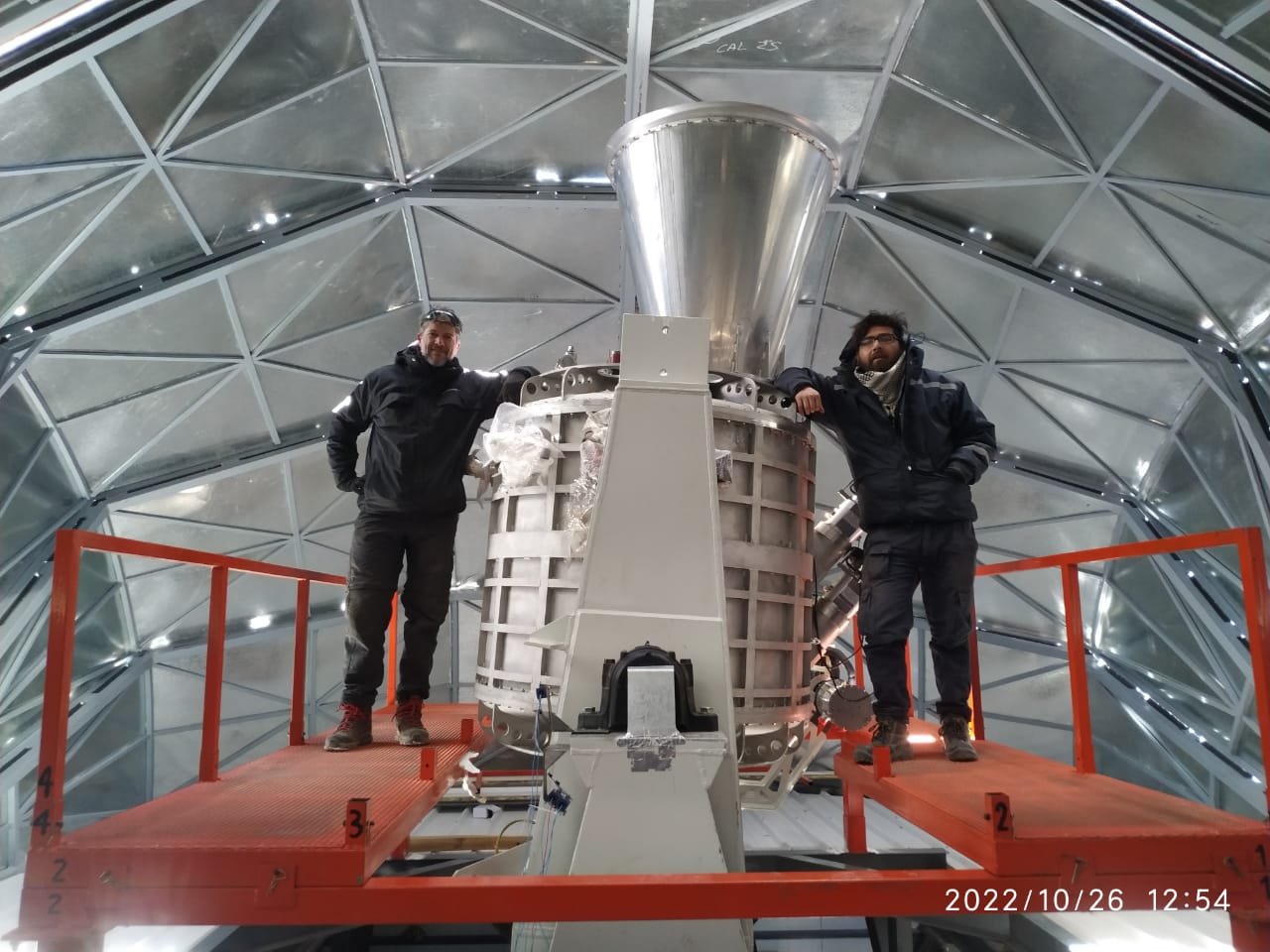}
    \caption{QUBIC instrument during assembly and integration on the observatory site (5000 m above sea level at Altos Chorrillos), featuring IAR's System Engineer Emiliano Rasztocky and ITEDA's technician Fabricio Rodriguez. Photo by Manuel Platino.}
    \label{fig:qubic}
    \end{figure}

\begin{figure}[!t]
    \includegraphics[width=\columnwidth]{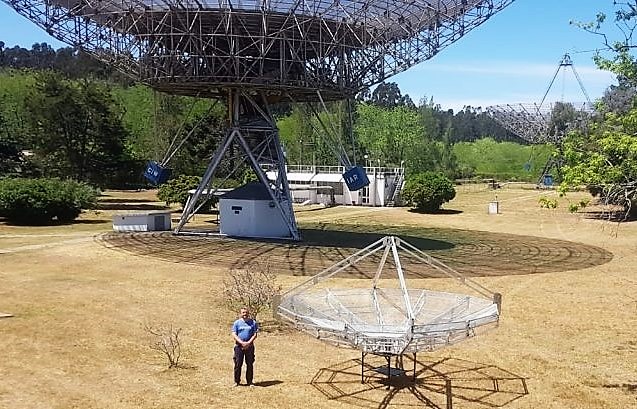}
    \caption{First model for the Multipurpose Interferometric Array standing next to IAR's main radio telescope Carlos M. Varsavsky. By the antenna, IAR's Lead Mechanical Technician, Pablo Alarcón.}
    \label{fig:mia_antena}
    \end{figure}

IAR's commitment to TT is further exemplified by its involvement in groundbreaking scientific projects such as MIA (Multipurpose Interferometer Array) and LLAMA (Large Latin American Millimeter Array). MIA focuses on designing and developing a first-of-its-kind radio telescope in South America, showcasing the institute's expertise in antenna design and development in novel applications for radio astronomy (see Fig. \ref{fig:mia_antena}). Additionally, the LLAMA project aims to create a state-of-the-art millimeter-wave observatory, fostering collaboration among Latin American countries. IAR is also contributing to the development of a ground station derived from the MIA project, enhancing communication and data reception capabilities for various small satellite missions. Another notable project in which the IAR is involved is the QUBIC project in the province of Salta, which aims to deploy a cosmic microwave background (CMB) observatory, contributing to our understanding of the early universe (see Fig. \ref{fig:qubic}).

The article is organized into seven distinctive sections, each delving into crucial aspects of the IAR's technology transfer endeavors.

In section two, we delve into the theoretical framework underpinning technology transfer and innovation in Latin America and on a global scale. The Sábato-Botana Triangle, the Helix Model, and Open Science and Innovation are explored as key models that have fostered collaboration and societal impact through TT.

Section three provides a context for the IAR's work in technology transfer, by discussing the National System of Science and Technology in Argentina. It also addresses the resilience displayed during the challenging times of the COVID-19 pandemic, by assessing and develop solutions during the health crisis in the country.

Section four offers an overview of the IAR's mission and vision for TT. It discusses the Institute's organizational transformation, timeline of technology transfer, and briefs ongoing projects spanning several industry sectors.

Section five discusses the IAR's working lines for technology transfer, which are focused on building strong relationships with SME’s, the scientific field, government agencies, and society at large.
Section six tackles the importance of understanding the needs and issues of SMEs, in order to effectively transfer knowledge and technology from the field of radio astronomy to these sectors.

Section seven envisions the Institute's strategies for its TT goals in the future, while strengthening collaborations at local and international levels and fostering Public-Private Partnerships (PPPs) with commerce and productive chambers. It also comprises the measures which must be taken to optimize TT processes and deliver effective solutions to promote innovation and collaboration.

Finally, it concludes with the IAR’s continuous efforts to bridge the gap between scientific research and practical applications. By actively engaging with industry partners, governmental agencies, and interdisciplinary collaborations, the IAR has demonstrated its capacity to transfer knowledge and technology from the field of radio astronomy to diverse sectors with socioeconomic impact.

% xxxxxxxxxxxxxxxxxxxxxxxxxxxxxxxxxxxxxxxxxxxxxxxxxxxxxxxxxxxxxxxxxxxxxxxxxx
\section{Technology Transfer and Innovation Models in Latin America and the Global Context}
\label{sec:TTmodels}
While it is beyond the scope of this work to provide a comprehensive overview of technology transfer and innovation models, it is worth mentioning some of the key models and authors that have influenced the field.

The study of technology transfer and innovation models has experienced significant growth over the past few decades \citep{LopezRubio2018,Wahab2012}. Researchers have extensively explored various models to understand the dynamics of transferring technology and knowledge across different contexts and industries. These models encompass a wide range of strategies and approaches for drive innovation and ensuring the effective dissemination of technology \citep{Ramanathan2008,Choi2009}.

\subsection{The Sábato-Botana Triangle}
\label{sec:sabatotriang}
From a local perspective, in the late 1960s, the emergence of the ``Latin American thought on science, technology, and society'' (PLACTS, Pensamiento Latino Americano en Ciencia, Tecnología y Sociedad) marked a turning point in the field. Authors such as Oscar Varsavsky \citep{Varsavsky1969}, brother of Carlos M. Varsavsky IAR's first director, Jorge Sábato, Natalio Botana, Amílcar Herrera, Jorge Katz, among several others, contributed significantly to this movement \citep{Dagnino1996,Carro2021,Feld2011}.

\begin{figure}[!t]
    \includegraphics[width=\columnwidth]{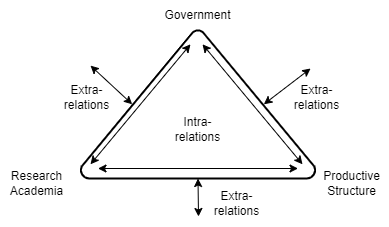}
    \caption{Sábato-Botana Triangle of relationships (see Sect. \ref{sec:sabatotriang}).}
    \label{fig:TTmodeltriangulo}
\end{figure}

A defining milestone in the field of technology linkage and innovation, particularly in the Latin American context, was the publication of the 1968 paper ``Science and Technology in the future development of Latin America" by Jorge Sábato and Natalio Botana \citep{Botana1970}. In their work, they analyzed the concept of the Triangle of Relationships between Government, Science-Technology, and Productive Structure, later known as the Sábato-Botana Triangle, which continues to exert considerable influence over the region's approach to technology appropriation, development and industrialization (see Fig. \ref{fig:TTmodeltriangulo}).

The Sábato-Botana Triangle is a model of economic development that emphasizes the importance of industrialization through import substitutions as a means to reduce dependence on foreign technologies. By promoting domestic innovation and self-sufficiency, governments seek to replace imported goods and technologies with locally produced alternatives, facilitating economic growth and stability.

In honor of Jorge Sábato's legacy and contributions, the 4th of June has been officially designated as the National Day for Technological Linkage since 2019. This day serves as a recognition of Sábato's ideas and as an occasion to highlight the importance of technology transfer as a driving force behind economic growth, societal development, and global competitiveness.

\subsection{The Helix Model}
\label{sec:helixmodel}
\begin{figure}[!t]
    \includegraphics[width=\columnwidth]{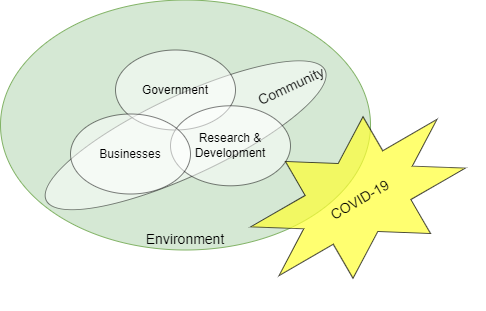}
    \caption{Quadruple helix concept by Etzkowitz and Leydesdorff (see Sect. \ref{sec:helixmodel}).}
    \label{fig:TTmodelhelix}
\end{figure}

The Triple-Helix model, proposed by Henry Etzkowitz and Loet Leydesdorff, is closely related to Sábato and Botana’s Triangle. This model may add a fourth helix to the original model, representing civil society and societal actors (see Fig. \ref{fig:TTmodelhelix}). The model, which is widely adopted today, recognizes the importance of societal engagement and social innovation in the process of technology development and solutions finding. It highlights the role of universities, industries, and governments, but also emphasizes the involvement of non-governmental organizations, community groups, and other stakeholders \citep{Leydesdorff1998}.

\subsection{Open Science and Innovation}
\label{sec:openinnovation}
\begin{figure*}[ht!]
    \centering
    \includegraphics[width=0.75\paperwidth]{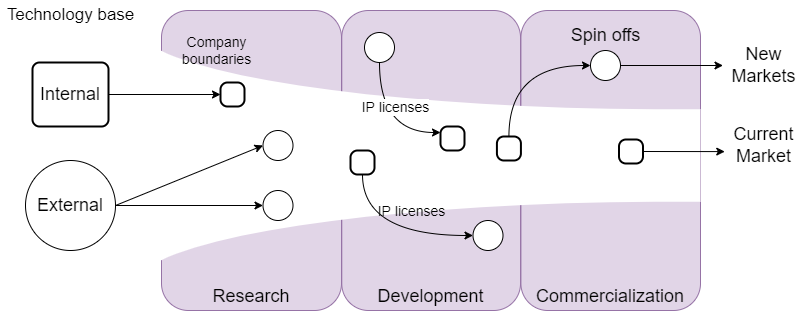}
    \caption{Open innovation model synthesis by \citep{Chesbrough2003}.}
    \label{fig:TTmodelopeni}
\end{figure*}

Another concept that has significantly influenced the field of technology transfer and innovation is Open Innovation. Coined by Henry Chesbrough, Open Innovation is a paradigm shift in how organizations approach the process of innovation \citep{Chesbrough2003}. Traditionally, innovation was seen as an internal process within a closed system, where companies relied on their own research and development (R\&D) to generate new ideas and bring them to market (see Fig. \ref{fig:TTmodelopeni}).

It also recognizes that valuable knowledge and ideas can come from both internal and external sources. It suggests that organizations should actively seek external knowledge, ideas, and technologies, and allow their own unused or underutilized knowledge to be used by others. This open and collaborative approach to innovation fosters partnerships and knowledge-sharing among various stakeholders, including universities, research institutions, industries, and even the public.

Open Science is a related concept that aligns with the principles of Open Innovation. It emphasizes making scientific research and data more accessible and transparent to a broader audience, including other researchers, policymakers, and the general public. It encourages the sharing of research findings, data-sets, methodologies, and even negative results, accelerating the pace of scientific discovery and technological development, some of which are related to the impact of Industry 4.0 \citep{Alkhazaleh2022,Stone2012}.

Both Open Innovation and Open Science can play vital roles among universities and research institutions, which can adopt these practices to disseminate their research findings more effectively, enabling industry partners to access and use relevant knowledge for commercial applications \citep{VicenteSaez2020}\footnote{See also ``Diagnosis and roadmap for an Open Science policy in Argentina'', Comité Asesor en Ciencia Abierta y Ciudadana 2022,  technical report, http://sedici.unlp.edu.ar/handle/10915/150523}. Similarly, industries can engage in an Open Innovation process by collaborating with external partners, including universities and startups, to leverage their expertise and technologies for developing new products and services, leading to the development of innovative solutions that address complex societal challenges and promote sustainable economic growth \citep{Audretsch2012,Harsanto2022,Morisson2022,Eppinger2021}.

\subsection{Looking Down to Earth: Technology Transfer in Radio Astronomy}
\label{sec:TTradioastronomy}
Radio astronomy stands out as a field with unique strengths, owing to its multidisciplinary nature. The technology and innovations crafted for radio astronomical research possess the potential for versatile applications across various industries and production domains \citep{Willebrands2020}.

Remarkable cases of technology transfer from radio astronomy have had a significant impact on the advancement of astronomy and space sciences in South African and Australian societies \citep{Povic2018,McBride2018}, as seen with the Square Kilometer Array (SKA), MeerKAT \citep{Jonas2018} and the LOw Frequency ARray (LOFAR) in the Netherlands \citep{Schakel2009}. This highlights the potential for developed regions to contribute to the scientific growth of emerging regions through technology transfer.

In the medical field, technologies initially designed for radio astronomy have found use in the development of medical devices and diagnostics, improving healthcare outcomes \citep{Hughes2007}. It also extends to electronics, digital design, and data processing, by employing an Open-Source paradigm as it happens with the technology developed by CASPER \citep{Hickish2016}. Additionally, the cross-disciplinary approach of radio astronomy has paved the way for novel techniques, like the case of detecting water leaks in pipes that are part of water networks in infrastructure systems \citep{Samtani2008}.

\subsection{Constraints and Impact}
\label{TTconstraints}
In the context of TT within the Science and Technology National System (SNCT, Servicio Nacional de Ciencia y Tecnología) and specifically for CONICET, there are various constraints that can arise. These constraints pose challenges and considerations when it comes to transferring scientific knowledge and technological innovations to practical applications \citep{Shmeleva2021,Audretsch2012,Nikitin2023}\footnote{See also ``Transfer of technology and knowledge sharing for development: science, technology and innovation issues for developing countries'', United Nations Digital Library, http://digitallibrary.un.org/record/784430}. Some of the key constraints include:

\begin{itemize}
    \item \textbf{High-risk investment}: Technology transfer often involves investments in R\&D, which can be high-risk endeavors. The uncertainty surrounding the success and market viability of new technologies can make it challenging to attract investment from private entities.
    \item \textbf{Unknown markets}: When transferring technology, there may be uncertainties regarding the market demand and potential adoption of the innovation. Understanding the target market and identifying potential customers or users can be complex, especially for cutting-edge technologies with limited market precedents.
    \item \textbf{Long-term perspective}: Technology transfer requires a long-term perspective, as the process of moving from R\&D to commercialization can be time-consuming. It often involves multiple stages, including prototyping, testing, regulatory approvals, and market validation, which require sustained efforts and resources.
    \item \textbf{Innovation-driven challenges}: Technological innovations are often disruptive and can face resistance from established industries or regulatory frameworks. Overcoming regulatory hurdles and addressing potential conflicts with existing technologies or business models can pose additional challenges.
    \item \textbf{Cognitive appropriation and blind technology transfer}: Research funded by public institutions, may face the challenge of ``cognitive appropriation" or knowledge leakage, where patentable inventions originating from public financing end up being owned predominantly by private and foreign actors. Another phenomenon, often referred to as ``blind technology transfer", involves 
    the scientific knowledge from developing countries, used to develop new technologies in foreign countries without the participation or benefit of the developing countries. Both cases can have a number of negative consequences, including the loss of potential economic benefits and the undermining of the development of indigenous technological capabilities \citep{Codner2012,Zukerfeld2022a,Zukerfeld2022}.
\end{itemize}

Despite these constraints, technology transfer initiatives, such as patents, licenses, startups, and in-house/on-demand research, continue to make a significant impact and have gained new momentum with the establishment of technological parks and incubators. Promotions within research centers like CONICET and Universities have also contributed to a remarkable surge in patent applications and grants. The number of patent applications has grown significantly, from 1 to 3 patents per year from 1970 until 1999, to 105 and 85 applications in 2017 and 2020, respectively\footnote{¿Qué se patenta en Argentina? Solicitudes de patentes presentadas por organismos de ciencia y tecnología, Dirección Nacional de Estudios, Subsecretaría de Estudios y Prospectivas, MinCyTI 2022, \url{https://www .argentina.gob.ar/sites/default/files/patentes\_oct.pdf}},\footnote{CONICET en cifras, 2022, \url{https://cifras.conicet.gov.ar/ publica/detalle-tags/26}} \citep[see also][]{Lusi2020}.

From the perspective of IAR, the following strategy is being followed to create a significant social and productive impact:
\begin{itemize}
    \item \textbf{Highly specialized consultancy services}: One of CONICET’s main assets is its capability, through its network of Institutes, to provide expertise and consultancy services in several disciplines and knowledge areas. In particular, IAR is able to offer the know-how in radio astronomy, antenna design, signal processing, and radio frequency (RF) related fields as valuable insights and solutions to SMEs partners and industry collaborators \citep{Bocconi2015}.

    \item \textbf{Networking within the Science and Technology System}: The SNCT -acronym that stands for Sistema Nacional de Ciencia y Tecnología- has access to a vast network of researchers, institutions, and organizations, enabling collaboration and knowledge exchange. This network facilitates the sharing of ideas, expertise, and resources, fostering a collaborative ecosystem for technology and knowledge transfer \citep{Vicente2022}.
    
    \item \textbf{Ready availability of services and infrastructure}: IAR's infrastructure and resources, including its laboratories, testing facilities, and specialized equipment, are readily available for technology transfer projects and services offer. This accessibility accelerates the development and validation of new technologies, reducing time-to-market needs from private companies.
    
    \item \textbf{SME focus}: IAR's technology transfer efforts are particularly focused on supporting SMEs. By providing access to expertise, resources, and collaborative opportunities, IAR helps SMEs overcome technological barriers and encourages their growth and competitiveness.

    \item \textbf{Limited time contracts}: IAR offers technology transfer services through limited time contracts, which allow efficient resource-allocations and collaborations. This approach ensures that projects are effectively executed within specific timeframes, enabling timely outcomes and results.
\end{itemize}

% xxxxxxxxxxxxxxxxxxxxxxxxxxxxxxxxxxxxxxxxxxxxxxxxxxxxxxxxxxxxxxxxxxxxxxxxxx
\section{Overview of the National System of Science and Technology}
\label{sec:overviewSNCT}
The SNCT in Argentina is a comprehensive network of organizations, institutions, and agencies that work together to promote scientific research, technological development, and innovation. It is essentially defined by the Ministry of Science, Technology, and Innovation (MinCyT), while the Consejo Interinstitucional de Ciencia y Tecnología (CICYT) 
is the main coordinating body of the SNCT\footnote{Resolución 319/08, MinCyTI 2008, \url{https://www. argentina.gob.ar/sites/default/files/resol\_mincyt\_319\_08.pdf}}. The CICYT's objective is to foster and improve relations among the scientific and technology member institutions, as stated in its founding law 25.467 from 2001\footnote{Ley 25.467, Sistema Nacional de Ciencia, Tecnología e Innovación, PEN 2021, \url{http://servicios.infoleg.gob.ar/infolegInternet/anexos/65000-69999/69045/norma.htm}}.

CONICET is one of the leading organizations within the SNCT, which accounts for almost a quarter of the total number of staff members in the SNCT, with more than 25,000 researchers and professionals, over 300 research institutions, and operates with a budget of USD 300 million, as of 2021\footnote{Indicadores de Ciencia y Tecnología Argentina 2021, Dirección Nacional de Información Científica, MinCyTI 2023, \url{https://www.argentina.gob.ar/sites/default/files/2018/05/ indicadores\_2021-web.pdf}}. 

From a technology transfer point of view, CONICET has been increasing its production of patents and licenses, with more than 1,100 patents granted between 1971 and 2021.
%[Comparación con el número de patentes de IBM y otras tecnologicas]
In recent years, CONICET has also been supporting the creation and promotion of more than 50 tech-based startups. Additionally, in 2021 alone, CONICET signed 3,500 agreements and offered at least 13,500 technological services to SMEs, both national and international, and other member institutions of the SNCT$^{5}$,\footnote{PLAN NACIONAL DE CIENCIA, TECNOLOGÍA E INNOVACIÓN 2030, Dirección Nacional de Políticas y Planificación, MinCyTI 2022, \url{https: //www.argentina.gob.ar/sites/default/files/plan\_nacional\_de\_ cti\_2030.pdf}}.

In addition to the aforementioned achievements, there have been several remarkable milestones and endeavors in the SNCT since the introduction of the law. These initiatives have significantly propelled scientific research, technological development, and innovation in Argentina. Some clear examples include:

\textbf{Mixed-ownership companies}:
\begin{itemize}
    \item \textit{INVAP}: Leading technological company that specializes in nuclear energy, complex medical devices, and satellites.
    \item \textit{ARSAT}: Government-owned geo-telecommunications company dedicated to telecommunications satellites and satellite television.
    \item \textit{Y-TEC}: A technology development company in the oil and gas sector, jointly owned by YPF (51\%) and CONICET (49\%).
    \item \textit{VENG}: Aerospace company devoted to manufacturing space launchers, ground station operation and technology development.
\end{itemize}

\textbf{Technological Parks and Business Incubators} \citep{Chaves2019}\footnote{See also ``Parques industriales, parques tecnológicos e incubadoras de empresas en Argentina'', La enciclopedia de ciencias y tecnologías en Argentina, 
ECYT-AR 2017, \url{https://cyt-ar.com.ar/cyt-ar/index.php?title=Parques\_industriales,\_parques\_tecnológicos\_e\_incubadoras\_de\_empresas\_en\_Argentina}}:

\begin{itemize}
    \item \textit{Parque Tecnológico Litoral del Centro SAPEM}: This technological park brings together the National Universities, CONICET, municipal and provincial governments of Santa Fe to provide opportunities for technology-based companies from pre-incubation until they establish themselves.
    \item \textit{Incubadora de empresas UNC}: This business incubator supports technology-based companies affiliated with the National University of Córdoba. It offers infrastructure, guidance, and training for new businesses.
    \item \textit{Incubacen}: An incubator for technology-based companies associated with the Faculty of Exact Sciences at the University of Buenos Aires (UBA).
    \item \textit{CITES}: Venture builder with a deep tech focus, investing in scientific-based startups at their formation stage. CITES co-creates with founders, providing resources, expertise, and guidance.
\end{itemize}

\subsection{Technology Transfer \& Innovation During the COVID-19 Pandemic in Argentina}
\label{TTcovid-19}
Following the World Health Organization's (WHO) declaration of COVID-19 as a pandemic in February 2020, governments worldwide confronted a myriad of issues and uncertainties, urging immediate and sustained efforts to tackle the crisis and its far-reaching impacts.

The multifaceted nature of the pandemic, encompassing health crises, financial aid, lockdowns, logistics, and more\footnote{Sectors and businesses facing COVID-19: Emergency and reactivation, CEPAL 2020, \url{https://hdl.handle.net/11362/45736}}
 \citep[see also][]{Acosta2020}, demanded a collaborative approach involving active interactions between governments, industries, research institutions, and society at large. This cooperative effort aimed to effectively deal with the ongoing problems imposed by the pandemic.

As a result, new solutions emerged based on the concepts of open science, open innovation, and the quadruple-helix (or N-helix) models \citep{Ibanez2021}. In developing countries like Argentina, a combination of different approaches was adopted. Some initiatives were directed by the government, such as the development of COVID-19 diagnostic detection kits (NEOKIT and COVIDAR IgG). Private companies like TECME, Leistung, and CEGENS, which manufacture medical devices such as mechanical respirators, collaborated with MinCyT and state-funded companies like INVAP and VENG to significantly increase their production capacity by 300 \%. Additionally, the textile company Kovi partnered with CONICET in a co-creation effort to produce a protective face mask called ATOM PROTECT. These examples exemplify the successful implementation of open innovation, the quadruple-helix model, and their alignment with the principles of the Sábato-Botana Triangle\footnote{Pequeñas y medianas empresas argentinas desarrollan junto al INTI soluciones innovadoras para enfrentar la pandemia, INTI 2020, \url{https://www.inti.gob.ar/assets/uploads/files/vinculacion-intitucional/05/05\_newsletter\_octubre\_2020\_ingles.pdf}},\footnote{La Argentina frente al COVID-19: desde las respuestas inmediatas hacia una estrategia de desarrollo de capacidades, Red ISPA 2020.}.

IAR also demonstrated its rapid response capabilities from day zero, mobilizing a dedicated team to devise solutions to address the COVID-19 crisis. The team's efforts were applied on transitioning the country-level lockdown and isolation measures while analyzing strategies to ease restrictions as the situation improved. Led by the Institute's Liaison and Technology Transfer Area (LTTA), a combination of open innovation and collaborative support was harnessed to confront the challenges posed by the pandemic.

\begin{figure}[!t]
    \includegraphics[width=\columnwidth]{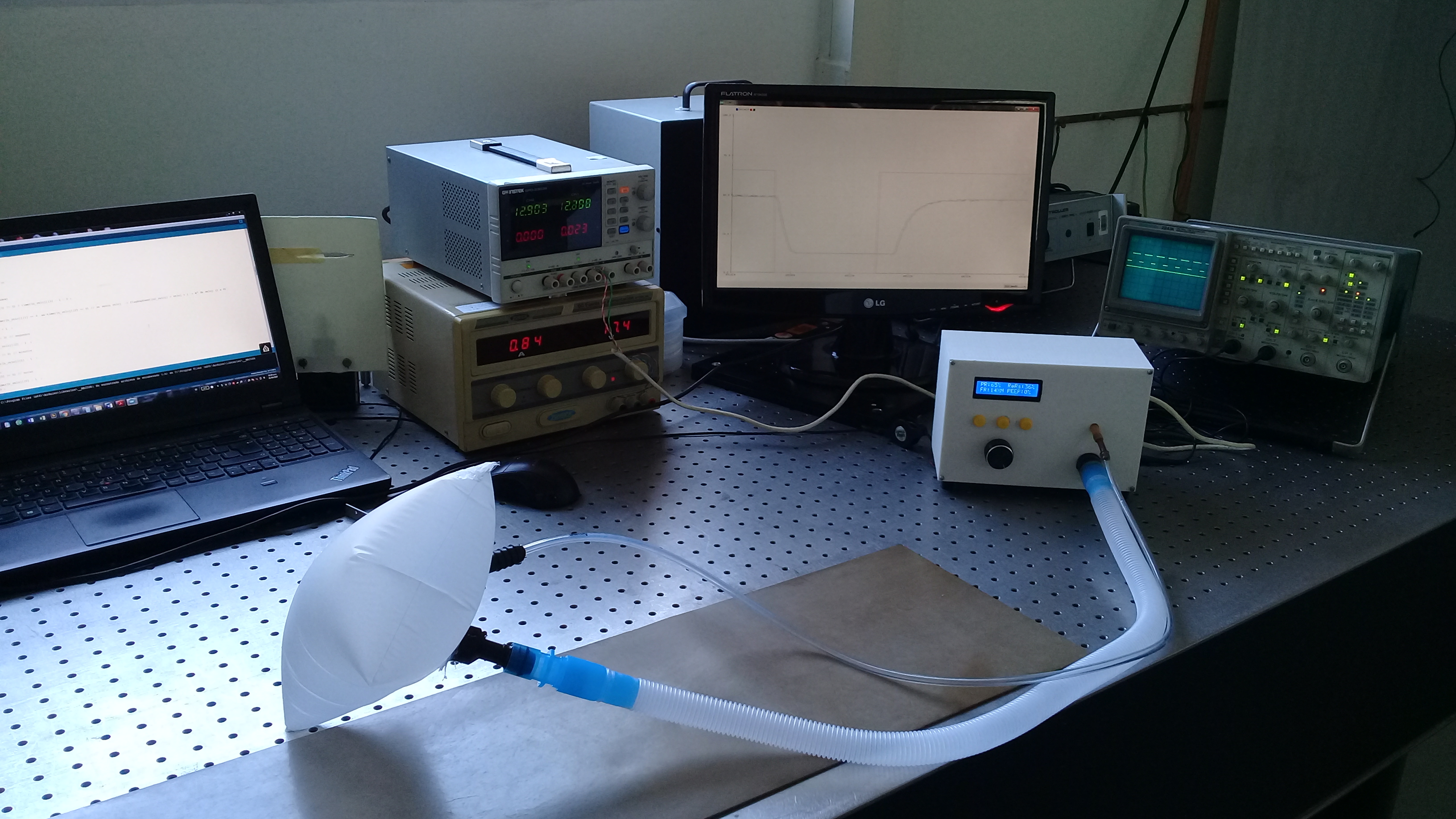}
    \caption{Prototype for the Non-Invasive Mechanical Ventilator for the COVID-19 pandemic.}
    \label{fig:covid-19vmni}
\end{figure}

\begin{figure}[!t]
    \centering
    \includegraphics[width=0.75\columnwidth]{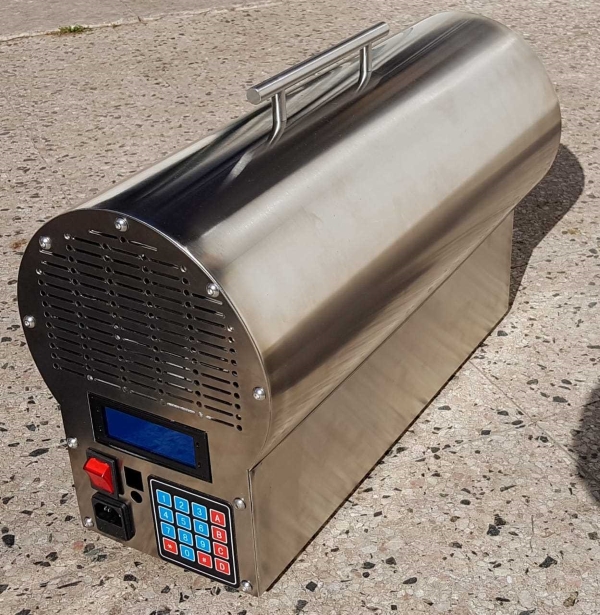}
    \caption{Production model of the first ozone reactor to be used in public space during the COVID-19 pandemic.}
    \label{fig:covid-19ozono}
\end{figure}

Among the first projects undertaken was the design and development of IARespira, a non-invasive mechanical ventilator (NIMV or NIV) tailored for the treatment of moderate COVID-19 patients during the outbreak \citep{Salibe2021} (see Fig. \ref{fig:covid-19vmni}). Collaborating with COVID-19 task force groups, the need arose to secure and rapidly open public spaces such as offices, classrooms, buses, trains, and emergency rooms. In response, IAR collaborated with a local SME, Acero a Medida S.A., to create an ozone generator for the destruction of viral loads in public environments \citep{Romero2020} (see Fig. \ref{fig:covid-19ozono}). Additionally, the Institute engaged in minor collaborations with hospitals and attended ministry-level discussions aimed at safely easing restrictions and reopening the economy at the earliest opportunity.

% xxxxxxxxxxxxxxxxxxxxxxxxxxxxxxxxxxxxxxxxxxxxxxxxxxxxxxxxxxxxxxxxxxxxxxxxxx
\section{IAR's Mission and Vision for Technology Transfer}
\label{sec:TTIARmyv}
IAR is constantly in pursuit of technological advancements and exploration of innovative projects of strategic significance. Central to its mission is the execution of cutting-edge technological initiatives, in which it has the potential to impact various industries beyond the sphere of radio astronomy. In the same way radio telescopes point towards the Universe, the technologies employed in observations and scientific research can also be directed towards enhancing people's activities and transforming businesses.

With a mission to \begin{displayquote}\textquote{develop and execute technological projects with great strategic value and foster cooperation between national and international institutions. Provide a training hub for new technologies and skills}\end{displayquote} the IAR is actively engaged in cultivating collaborative partnerships. Through these partnerships, knowledge exchange and technology transfer are facilitated, empowering the IAR to tap into a vast array of expertise and resources. This synergy enables the acceleration of technology development and cultivates an environment of shared growth.

Simultaneously, the IAR's vision is to \begin{displayquote}\textquote{become a benchmark institution in outstanding technological projects within the scientific and technological system. We aim to create a positive social impact to promote the development of our country.}\end{displayquote} By aligning its technological pursuits with its objective, the IAR is committed to making a meaningful impact on society, by addressing real-world challenges and catalyzing socio-economic development.

\subsection{Organizational transformation}
\label{sec:IARorganization}
The Institute has established a dynamic and forward-thinking approach to accelerate its organizational transformation and facilitate seamless collaboration between all actors involved. This strategic arrangement is designed to optimize project-defined resources with a tight cooperation between all levels, ensuring an efficient and agile process.

\begin{figure}[!t]
    \includegraphics[width=\columnwidth]{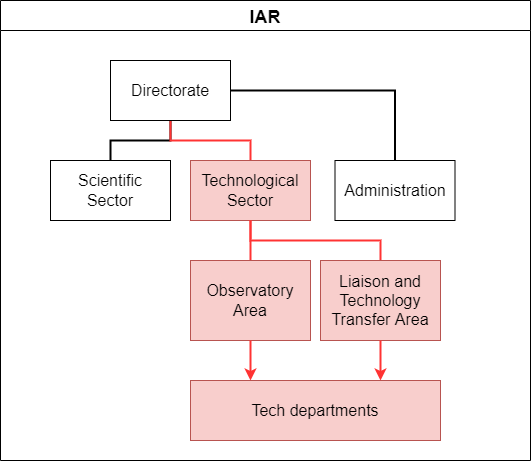}
    \caption{Organizational chart at IAR from a technological standpoint.}
    \label{fig:iar-organization}
\end{figure}
\begin{figure}[!t]
    \includegraphics[width=\columnwidth]{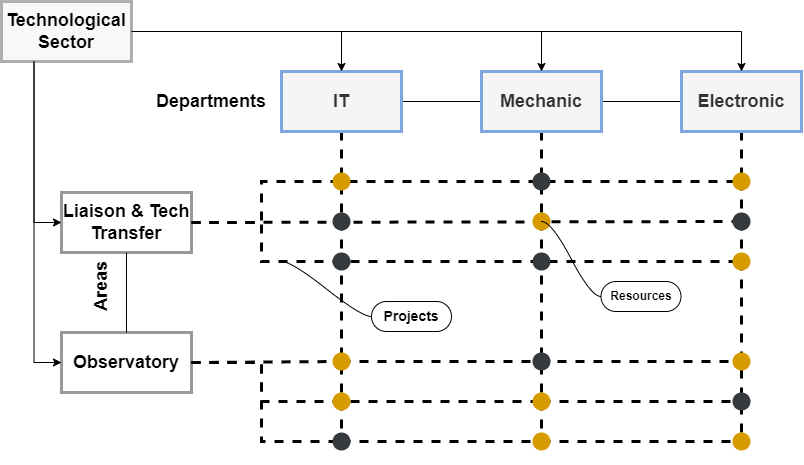}
    \caption{Matrix organization of IAR's Technological Sector.}
    \label{fig:iar-matrix}
\end{figure}
The organizational transformation began in 2018 with the definition of clear strategic and operational goals in its 4-year plan. The plan includes the establishment of new management areas, operational functions, and roles, affecting all levels of the organization.

The transformation is also leading to a higher qualitative level, as the staff is being trained in new skills and the processes are being streamlined to improve efficiency. The goal of the organizational transformation is to make the Institute more agile and responsive to the needs of social and economic demand.

Currently, the LTTA is working in close coordination with other areas of the IAR, which is helping to ensure that the Institute's technology transfer activities are aligned with its overall mission and vision.

At the helm of the IAR stands the Directorate, overseeing and coordinating the activities of the various sectors. The Administration sector handles the administrative aspects, providing support and resources to ensure the smooth functioning of the entire organization.

The IAR's Scientific and Technological Sectors constitute the core management pillars, in line with the Institute's foundational Act in the field of radio astronomy. These sectors are dedicated to advancing scientific research, establishing fruitful collaborations with fellow institutions, promoting education in radio astronomy, and nurturing the next generation of researchers, engineers, and technicians in their specialized domains. Additionally, they maintain vital scientific and technological relationships with national and international institutions\footnote{Foundation Act of IAR, 1966, \url{https://www.iar.unlp.edu.ar/institucional/finalidad/}}.
%\citep{IAR1966}.

The Technological Sector operates through two distinct and complementary operational areas: Observatory and Liaison and Technology Transfer (see Fig. \ref{fig:iar-organization}). Both areas are equipped with highly technical expertise, enabling a range of activities from observational studies, maintenance, and upgrades to new technological developments. Additionally, the sector embraces a commercial focus by providing services to local SMEs, stimulating relations between the SNCT and industry.

Firstly, the Observatory Area at IAR is equipped with two 30-meter radio telescopes. The antenna Carlos M. Varsavsky, operational since 1966, conducts extensive sky mapping of the southern hemisphere, while the Esteban Bajaja radio telescope inaugurated in 1977, focuses on radio continuum and polarimetric studies. Continuous upgrades enable the observation of phenomena like neutron stars and pulsar timing, aiding gravitational wave detection.

This area possesses the technical capacity to develop equipment necessary for detecting and receiving radio emissions from astronomical sources. This capability allows both radio telescopes to operate remotely and continuously, facilitating the study of diverse astronomical objects.
    
Here are some of the notable observational projects carried out or in progress at the IAR:

\begin{itemize}
    \item \textbf{PuMA (Pulsar Monitoring in Argentina)}: Involves monitoring the timing of known pulsars, particularly millisecond pulsars, in collaboration with the NANOGrav project for long-period gravitational wave searches. External financial support for this project is provided by Dr. Carlos O.  Lousto from the Rochester Institute of Technology\footnote{Integrated Station for Remote Pulsar Observations, Data Mining and Storage, RIT, Center for Computational Relativity \& Gravitation 2023, \url{https://ccrg.rit.edu/research/area/integrated-station} \url{-remote-pulsar-observations-data-mining-and-storage}}.
    
    \item \textbf{PUGLI-S (PUlsar Glitching Squad)}: Aims to intensively monitor bright pulsars with known timing jumps in the southern hemisphere.
    
    \item \textbf{Back-End for Deep Space Stations (DSA3 and CLTC-CONAE-NEUQUEN)}: Under a collaboration agreement with CONAE, the IAR is developing a digital Back-End for radio astronomical research. This instrument is based on CASPER hardware platforms \citep{Colazo2020} (see Fig. \ref{fig:obs-dsa}).
    
    \begin{figure}[!t]
    \includegraphics[width=\columnwidth]{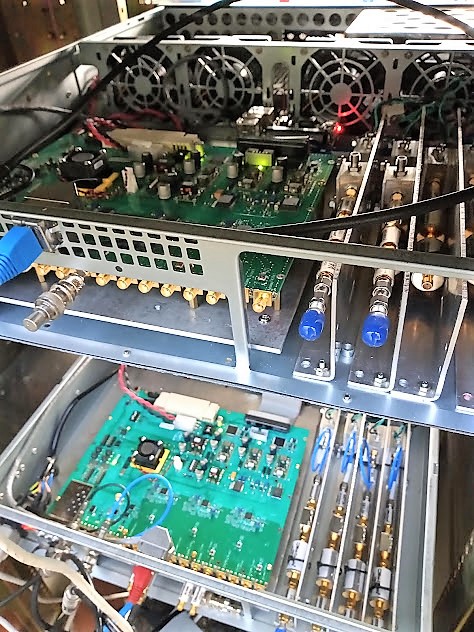}
    \caption{Digital Back-End (DBE) for the Deep Space Antennas.}
    \label{fig:obs-dsa}
    \end{figure}
    
    \item \textbf{Next Generation Event Horizon Telescope (ngEHT)}: As part of the next generation EHT project, the IAR collaborates in the development of a digital Back-End with a local tech-based company, CognitionBI (see Fig. \ref{fig:obs-eht}).

    \begin{figure}[!t]
    \includegraphics[width=\columnwidth]{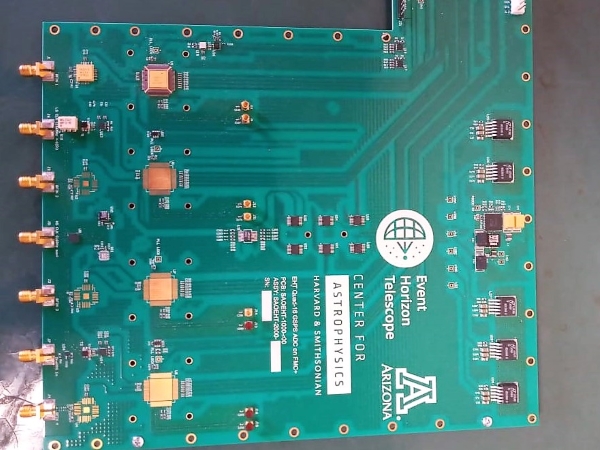}
    \caption{Digital Back-End (DBE) prototype for the ngEHT.}
    \label{fig:obs-eht}
    \end{figure}
    
    \item \textbf{Multipurpose Interferometric Array (MIA)}: The MIA project aims to create a multipurpose interferometer operating at low frequencies (50 MHz to 2 GHz). The array will consist mainly of five-meter-diameter antennas, enabling precise angular resolution and competitive scientific performance in Latin America.
    
    \item \textbf{Large Latin American Millimeter Array (LLAMA)}: Is a joint scientific and technological endeavor between Argentina and Brazil to establish and operate an instrument for astronomical observations in millimeter and submillimeter wavelengths. The collaboration extends internationally, including the Netherlands Research School for Astronomy (NOVA) \citep{Arnal2009} (see Fig. \ref{fig:obs-llama}).
    
    \begin{figure}[!t]
    \centering
    \includegraphics[width=0.75\columnwidth]{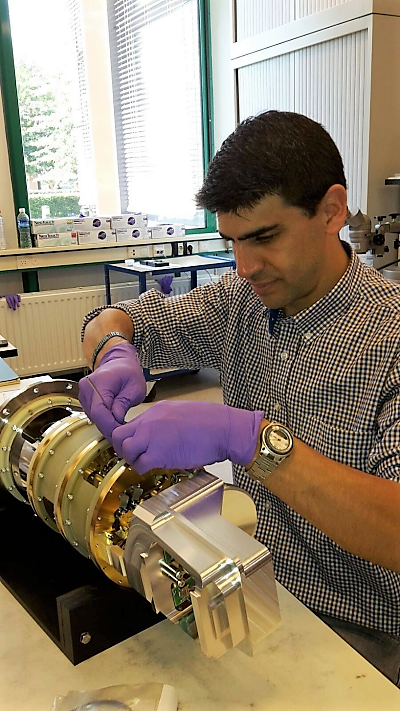}
    \caption{Guillermo Gancio during the integration of the LLAMA receptor at NOVA.}
    \label{fig:obs-llama}
    \end{figure}
    
    \item \textbf{Educational Outreach Projects}: The IAR engages in outreach activities with educational institutions, offering technical and scientific opportunities for students at the secondary and university levels. Students gain practical experience and conduct research related to various radio astronomy topics.
\end{itemize}

Secondly, the LTTA plays a critical role in connecting the IAR with external partners, including national and international institutions, industries, and other stakeholders. This area actively seeks opportunities for technology transfer and collaborative projects, encouraging synergistic relationships to translate scientific and technological developments into practical applications.

Finally, the Technological Sector operates with a matrix-style management approach, between the IT, Electro/Mechanical, and Electronics departments and the two operational areas (see Fig. \ref{fig:iar-matrix}). These technical units provide essential support and services, aligning their efforts with the Institution's needs and priorities.
The sector comprises a dedicated team of 33 members, with 17 skilled technicians and 11 highly qualified engineers. Their collective expertise allows the successful implementation of technology transfer and observational projects.

    \begin{figure*}[ht!]
    \centering
    \includegraphics[width=0.75\paperwidth]{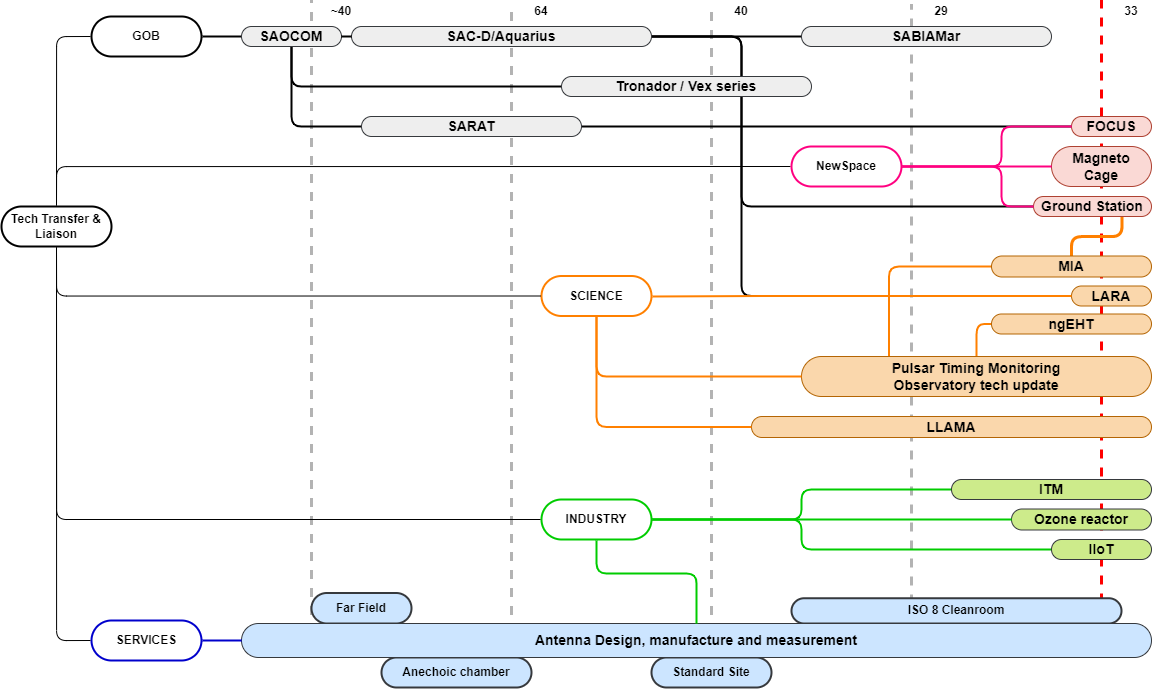}
    \caption{Timeline of IAR's technology transfer projects and activities over the last 25 years.}
    \label{fig:tt-time}
    \end{figure*}
    
\subsection{IAR’s Timeline of Technology Transfer}
\label{sec:TTtimeline}
Building upon the organizational transformation discussed earlier, the LTTA at the IAR has undergone several transformative stages over the past 25 years. This section traces the timeline of significant events and milestones that have shaped the IAR's technology transfer journey, emphasizing its dedication to collaboration and technology development across diverse sectors (see Fig. \ref{fig:tt-time}).

\begin{itemize}
    \item{\textbf{Early 2000s: the beginnings.}}
    The early 2000s marked a crucial phase for Argentina, characterized by a convulsed political and economic reorganization following the severe crisis of 2001. During this period, the Institute took proactive steps to address the nation's productive growth needs, responding to the renewed interest in science and technology.
    
    In the wake of the crisis, the country recognized the importance of investing in research, innovation, and technology transfer to overcome the scientific and technological debt. The IAR began offering its capabilities to meet these demands of external actors. These early services were in their nascent stage, yet they played a crucial role in laying the foundation for future technology transfer initiatives.
    
    \item{\textbf{2005 to 2015: formalization and complex projects.}}
    The IAR recognized the need to formalize its technology transfer initiatives to engage effectively with external partners. A gradual structured approach was set, beginning in 2005 and continuing through 2015.
    
    During this period, the IAR took on complex and demanding projects, collaborating with institutions such as CONAE, INVAP, VENG, and various space-related public-private partners. These developments held significant strategic value, propelling the Institute's expertise to new heights.
    
    The IAR became a key player in contributing to Argentina's earth observation satellite missions, SAC-D/Aquarius, SAOCOM, and SABIAMar. These missions showcased the Institute's capabilities and the dedication of its committed professional team. The opportunity to engage in such transformative projects allowed the staff to enhance their expertise, professionalize their approach, and continuously improve the Institute's facilities.
    
    As the IAR's reputation grew, a diverse array of SMEs expressed interest in its capabilities and services. These new partnerships demonstrated the broader impact of the IAR's technology transfer endeavors, reaching and engaging with emerging industries. As such, these initiatives solidify its position in technology transfer within the SNCT, with consistent results and a growing social impact.
    
    \item{\textbf{2018 to present: consolidation and expansion.}}
    Since 2018, the IAR has entered a consolidation and expansion phase in its technology transfer program, with new objectives and organizational structures to amplify its impact.
    
    Three primary drivers have shaped the current LTTA and its objectives. Firstly, a shift in political appreciation towards science and technology, along with the realization of the higher costs associated with the space ambitions of the country, prompted the need for a more cost-effective approach. This situation led to the search for new partnerships in the local NewSpace sector, while also intensifying technology transfer activities in the industrial domain.
    
    Secondly, with the appointment of a new IAR director in 2018, a vision was set to position the Institute on the international arena and bolster its influence at the national level. These measures facilitated an increased collaboration with new institutions, agencies, and SMEs.
    
    Lastly, the surge of the COVID-19 crisis required a rapid response from the IAR, leading to the development of effective solutions. Initiatives like a non-invasive ventilator and an ozone reactor to disinfect public spaces were undertaken, along with collaborations with hospitals, ministry-level assessments, and civil actors facing challenges during isolation and lockdown measures.
    
    The contributions made to the COVID-19 crisis, along with the reorganization and vision at the IAR, led to new partnerships which revitalized the technology transfer activities in various sectors, including education, medicine, IIoT, scientific collaborations (ngEHT and MIA pathfinder), and relationships with startups. 
\end{itemize}

\subsection{Infrastructure and resources available at IAR}
\label{sec:TTinfrastructure}
The IAR possesses diverse resources that facilitate technology transfer and the development of technological projects. This infrastructure cover an extensive area of more than 60,000 square meters, with a built surface of 1,400 square meters. Some of the key facilities include:

\begin{itemize}
    \item \textbf{Electronics Lab and Mechanical Workshop}: The electronics lab and mechanical workshop are equipped with advanced tools and equipment to design, build, and test electronic circuits and mechanical components. These facilities are vital for the development and prototyping of various projects, enabling the production of custom-made electronic devices and mechanical structures to meet specific requirements (see Fig. \ref{fig:tt-mecanica}).

    \begin{figure}[!t]
    \includegraphics[width=\columnwidth]{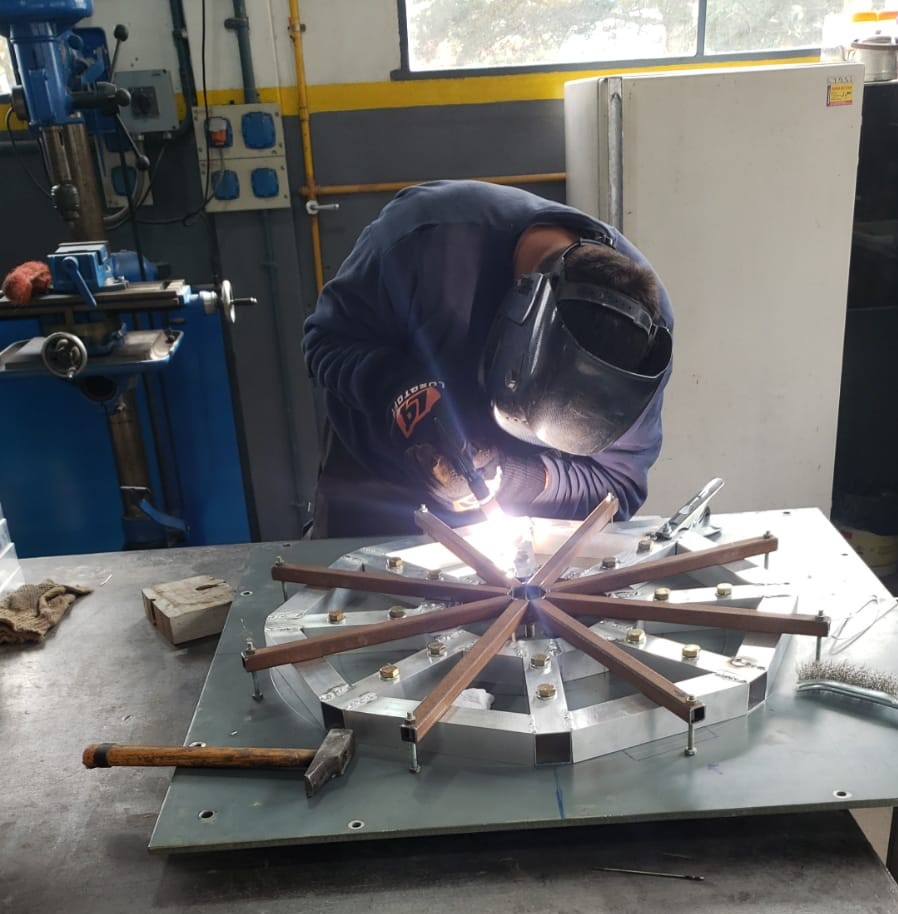}
    \caption{Technician Nahuel Duarte soldering the aluminium hub for the MIA antenna assembly in the mechanical workshop.}
    \label{fig:tt-mecanica}
    \end{figure}
    
    \item \textbf{ISO-8 Clean Room}: IAR houses an ISO-8 clean room with an ISO-5 laminar flow cabinet, with an overall area of 32 square meters. They provide a controlled environment with low levels of airborne particles, essential for the assembly and testing of sensitive electronic and optical components. It is crucial for manufacturing and assembling precision instruments and other critical technology transfer projects (see Fig. \ref{fig:tt-cleanroom}).

    \begin{figure}[!t]
    \includegraphics[width=\columnwidth]{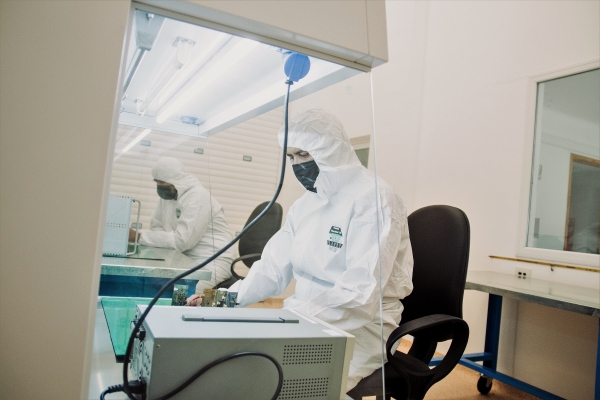}
    \caption{ISO-8 clean room during electronic assembly.}
    \label{fig:tt-cleanroom}
    \end{figure}
    
    \item \textbf{Far Field Range}: The FFR is an outdoor facility designed for measuring radiation patterns, cross-polar discrimination, antenna absolute gain, and radiation efficiency of antennas. It is a large open area of 120 meters in length and 9.5 meters in height. The range covers a frequency range of 4 MHz to 6 GHz and can handle antennas with a maximum size of 1 meter at 6 GHz. Its main use is characterizing and optimizing the performance of antennas and other communication systems, specially for satellite missions (see Fig. \ref{fig:tt-ffr}).

    \begin{figure}[!t]
    \includegraphics[width=\columnwidth]{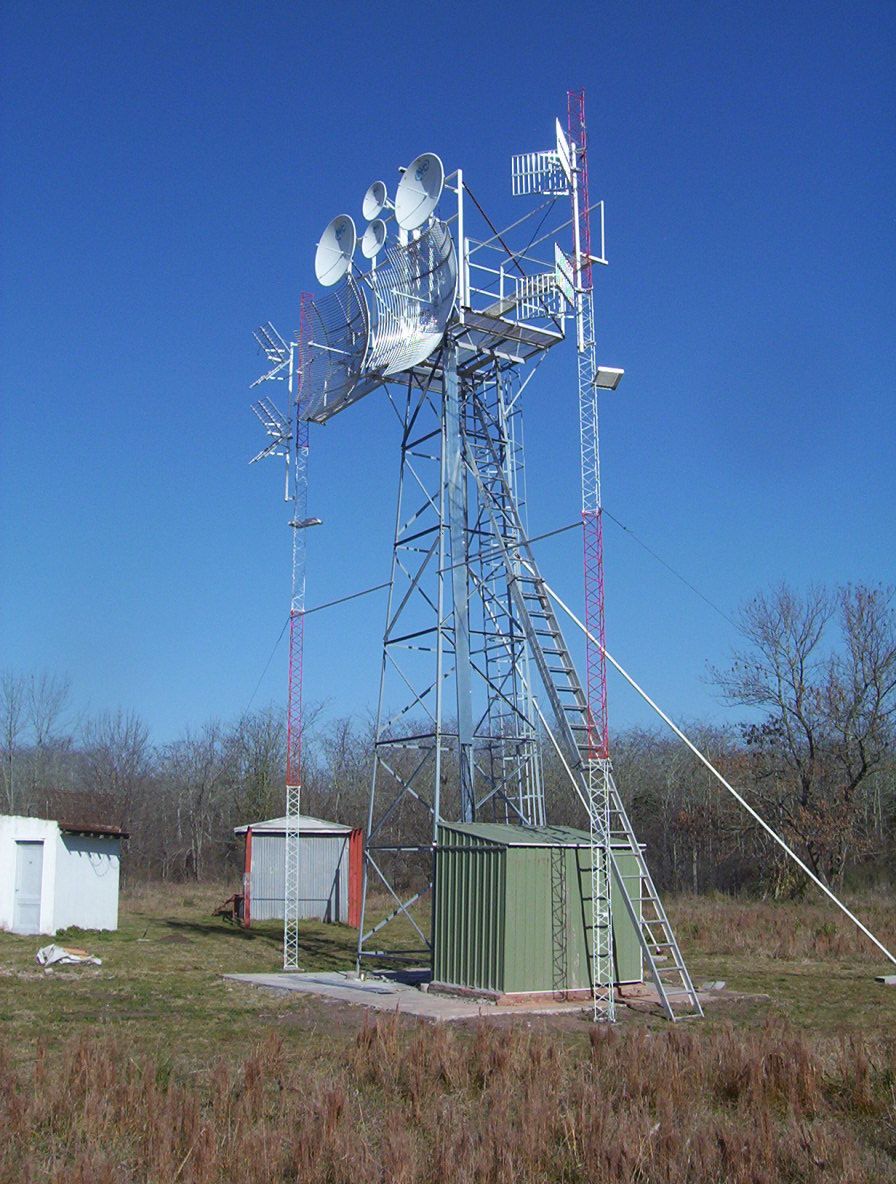}
    \caption{FFR antennas facilities.}
    \label{fig:tt-ffr}
    \end{figure}
    
    \item \textbf{Standard Site Area}: The SSA is another outdoor facility that measures cross-polar discrimination, antenna absolute gain, and antenna factor. It consists of multiple antennas placed at a distance of 10 meters from each other. The frequency range covered by this facility is from 500 KHz to 1 GHz, and it can accommodate antennas with a maximum size of 1.4 meters.
    
    \item \textbf{Anechoic Chamber}: It is a controlled electromagnetic environment that measures radiation patterns, cross-polar discrimination, antenna absolute gain, and radiation efficiency over the far field. It is a shielded enclosure of 12 meters in length, 8 meters in width, and 6 meters in height. Operates in a frequency range of 300 MHz to 40 GHz and can handle antennas with a maximum load of 200 kg. It also includes reference antennas in L, S, C, and X bands for calibration and comparison purposes (see Fig. \ref{fig:tt-cai}).
    
    \begin{figure}[!t]
    \includegraphics[width=\columnwidth]{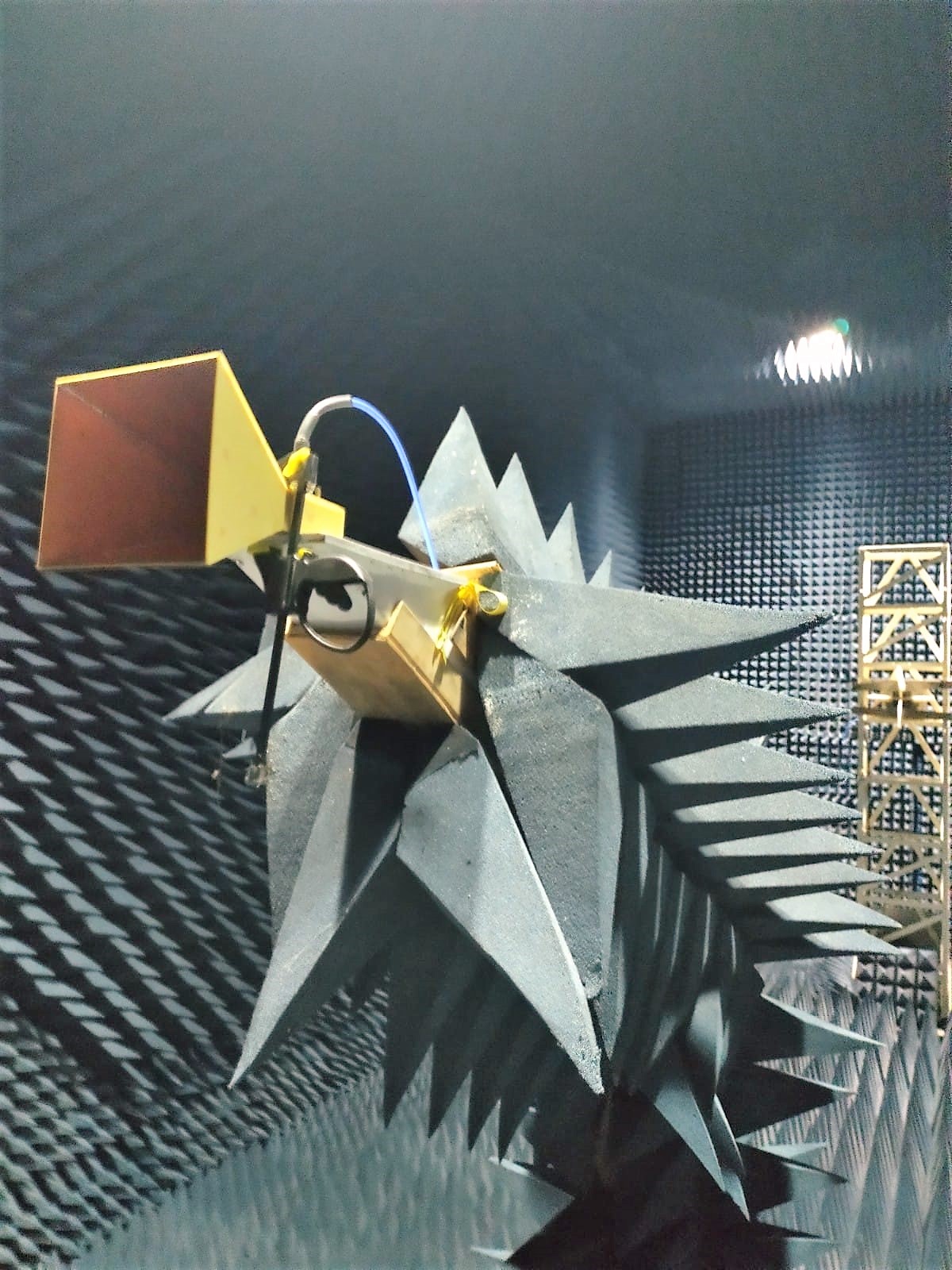}
    \caption{Antenna characterization tests inside the anechoic chamber.}
    \label{fig:tt-cai}
    \end{figure}
    
    \item \textbf{Cryo-Vacuum Lab}: Provides a controlled environment for conducting tests and experiments at cryogenic temperatures and vacuum conditions. This facility is essential for the development and characterization of instruments and components used in radio astronomy, like Low Noise Amplifiers (LNA) and equipment which require cryogenically cooled devices (see Fig. \ref{fig:tt-crio}).

    \begin{figure}[!t]
    \includegraphics[width=\columnwidth]{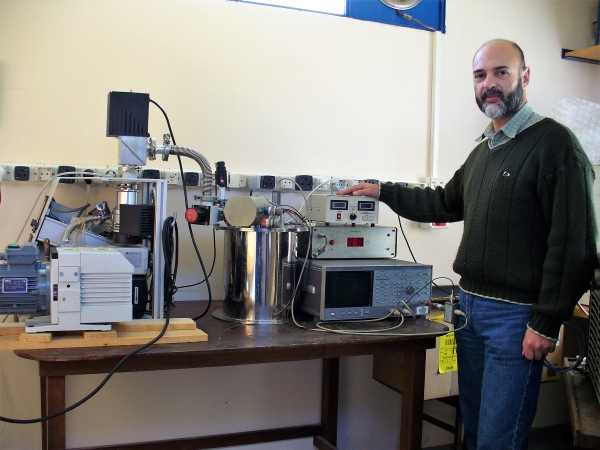}
    \caption{Ing. Daniel Perilli standing next to the Dewars cryostat.}
    \label{fig:tt-crio}
    \end{figure}
    
    \item \textbf{Antenna Design and Simulation Lab}: The Antenna Design and Simulation Lab is equipped with advanced software and simulation tools for designing and optimizing antennas. This lab plays a crucial role in the development of custom antennas tailored for specific applications, such as radio astronomy, satellite communication systems, radar technologies and IoT devices, among others.
\end{itemize}

\subsection{Ongoing Technology Transfer Projects}
\label{sec:TTprojects}
IAR is actively involved in a diverse range of technology transfer projects across various industries, including industrial applications, medical devices, and NewSpace initiatives.

\subsubsection{Industrial Applications}
\begin{figure}[!t]
    \includegraphics[width=\columnwidth]{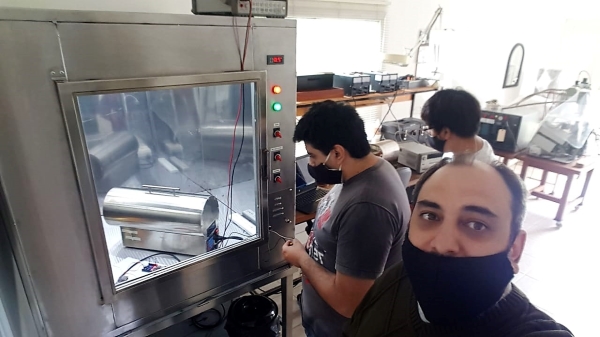}
    \caption{Martín Salibe, Gastón Valdéz and Luis Ferrufino testing the ozone reactor in a custom made calibration chamber.}
    \label{fig:tt-o3}
    \end{figure}

\textbf{COVID-19 Rapid Response - Automatic Ozone Reactor:} In response to the COVID-19 pandemic, IAR, together with CONICET and local SME Acero a Medida S.A., has developed an Automatic Ozone Reactor for disinfecting public spaces. The project is currently in its final stage of characterization and is being prepared for commercial use pending approval from regulatory agencies (see Fig. \ref{fig:tt-o3}).

\begin{figure}[!t]
    \centering
        \includegraphics[width=\columnwidth]{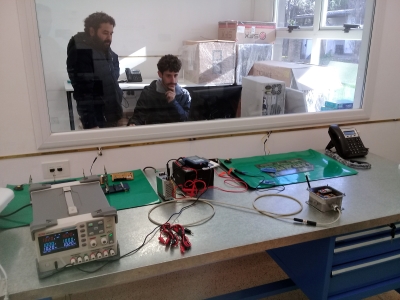}
        \includegraphics[width=\columnwidth]{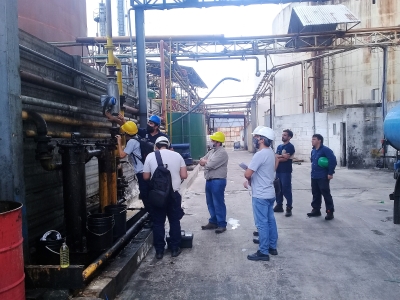}
        \caption{Upper panel: Matías Contreras and Augusto Donantueno analyzing data from the IoT nodes. Lower panel: CALM technical staff explaining the process to IAR's team.}
        \label{fig:tt-calm}
    \end{figure}

\textbf{IIoT for Remote Process Monitoring:} Is a collaborative effort involving the Federal Council for Science and Technology (COFECyT), under the Ministry of Science, Technology and Innovation; the worker-owned cooperative Aceitera La Matanza (CALM), and IAR. It comprises a network of Low-Cost IIoT Wi-Fi nodes, which is being designed, developed, and installed for remote monitoring of a sunflower oil refinery. Beyond the technology deployment, the project includes a comprehensive training program for users and technical staff, empowering them to utilize the new technology effectively (see Fig. \ref{fig:tt-calm}).

\subsubsection{Medical Devices}
\begin{figure}[!t]
    \includegraphics[width=\columnwidth]{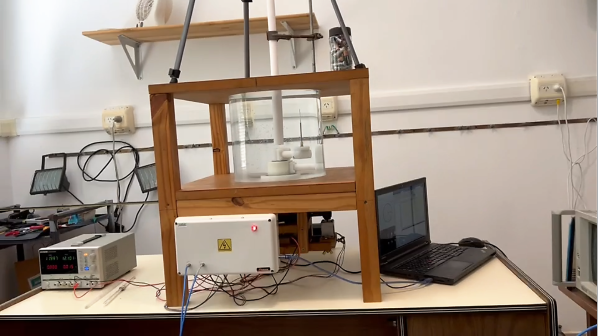}
    \caption{Data acquisition and validation tests with the ITM prototype.}
    \label{fig:tt-itm}
    \end{figure}

\textbf{ITM Image-based Microwave Tomography:} IAR, together with the Institute of Physics of Liquids and Biological Systems (IFLYSIB) and the UNLP, is working on a prototype stage of an Image-based Microwave Tomography (ITM). The primary objective of this project is to non-invasively and non-ionizingly measure changes in the dielectric characteristics of bone tissue. By processing signals transmitted and received with rotating monopoles in glycerin, the system aims to generate detailed images for medical diagnostics (see Fig. \ref{fig:tt-itm}).

\subsubsection{NewSpace}
\begin{figure}[!t]
    \centering
    \includegraphics[width=0.75\columnwidth]{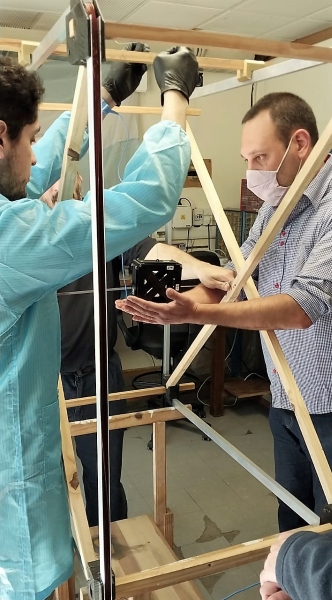}
    \caption{Elias Fliger performing V\&V tests with a 1-axis Helmholtz pair of coils.}
    \label{fig:tt-mdqsat}
    \end{figure}

\textbf{SmallSats V\&V:} Collaborating with Mar del Plata-based startup InnovaSpace, IAR is providing the development of a Helmholtz cage and test services for SmallSat verification and validation. The project involves testing satellite magnetotorquers and verifying the MDQSAT-1 Attitude Determination and Control System (ADCS) module's detumbling mode (see Fig. \ref{fig:tt-mdqsat}).

\begin{figure}[!t]
    \includegraphics[width=\columnwidth]{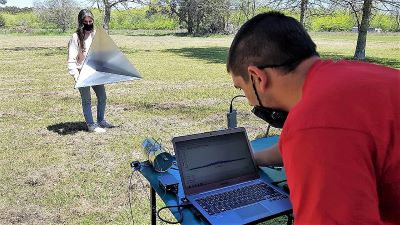}
    \caption{UNSAM students performing tests with FOCUS first prototypes.}
    \label{fig:tt-focus}
    \end{figure}

\textbf{FOCUS Project - X-band SAR Antenna Design:} Engaging in a Public-Private cooperation, IAR, along with CONAE, SME SpaceSur, and the National University of San Martin (UNSAM), is undertaking the design of an X-band Synthetic Aperture Radar (SAR) antenna proof of concept. The project involves an iterative design process, starting from a minimum viable product (MVP) in C-band and progressing to the Engineering Model (EM) in X-band. The ultimate goal is to achieve a 10 GHz irradiating panel with processing units, enabling improved remote sensing capabilities over existing cases (see Fig. \ref{fig:tt-focus}).

\textbf{Lunar Antenna for Radio Astronomy (LARA):} In collaboration with CONAE, IAR is working on a development proposal for the LARA instrument. Leveraging knowledge and experience from previous missions, including SAC-D/Aquarius and SABIAMar, the project is planning to deploy a payload on the L2 orbit. Inspired by the Netherlands-China Low-Frequency Explorer (NCLE) and CubeRRT, the design will operate in the frequency range of 30 MHz to 300 MHz. LARA aims to measure synchrotron radiation, Jupiter X-ray flares, solar bursts, and facilitate the detection of the Vela pulsar at low frequencies while addressing Radio Frequency Interference (RFI) challenges.

\begin{figure}[!t]
    \includegraphics[width=\columnwidth]{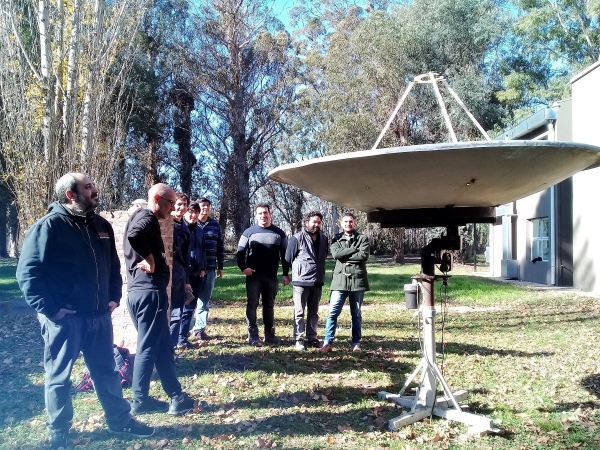}
    \caption{Martín Salibe, Leandro García, Santiago Spagnolo, Luis Ferrufino, Juan M. González, Matías Contreras and Darío Capucchio after the Ground Station Prototype installation.}
    \label{fig:tt-pet}
    \end{figure}

\textbf{Ground Station Prototype (PET):} As a spin-off from the Multipurpose Interferometric Array (MIA) project, IAR is developing a Ground Station Prototype focused on supporting smallsats. It also leverages existing engineering efforts from MIA, with the main goal to operate in the S-band, X-Band, and UHF frequencies, although it is still under discussion (see Fig. \ref{fig:tt-pet}).

% xxxxxxxxxxxxxxxxxxxxxxxxxxxxxxxxxxxxxxxxxxxxxxxxxxxxxxxxxxxxxxxxxxxxxxxxxx
\section{Working Lines for Technology Transfer}
\label{sec:TTlines}
The Institute pursues technology transfer through four distinct working lines, each tailored to address specific needs and requirements of its collaborative partners and stakeholders:
\begin{itemize}
    \item{\textbf{Project Agreements.}} Project agreements involve long-term contracts lasting more than six months and typically exceeding 2500 engineering hours. These agreements entail comprehensive technological developments and collaborations with various entities, including government agencies and organizations. Some examples include collaboration with the National Space Plan and CONAE since the early 2000s. These collaborations have been instrumental in the development of satellite missions such as SAC-D/Aquarius, SAOCOM, and SABIAMar, leveraging IAR's expertise in RF, telecommunications, analogue and digital electronics, firmware, and software design\footnote{Plan Espacial Nacional, CONAE 2016, \url{https://www.argentina.gob.ar/ciencia/conae/plan-espacial}} \citep{Lopez2018,Lopez2021,Alvarez2021}. The successful execution of these project agreements has significantly contributed to advancing space technology and strengthening Argentina's capabilities in space exploration.
    
    \item {\textbf{Consultancy.}} IAR offers consultancy as a technological service, providing expert assessment and design solutions across a wide range of topics, including RF, telecommunications, analogue and digital electronics, firmware, and software. This sort of contracts typically span periods below six months and involve less than 1000 engineering hours. This working line allows external organizations and industries to tap into IAR's expertise and related technologies to address specific challenges and optimize their systems.
    
    \item{\textbf{Test and Measurement (T\&M).}} IAR leverages its testing facilities, to offer test and measurement services. These services are provided on a contract basis per work package, offering external organizations access to a wide range of technologies and capabilities. Cases of T\&M projects include antenna design and characterization for ENACOM (Argentina's telecom regulatory agency), DirecTV, and COPITEC (Consejo Profesional de Ingeniería de Telecomunicaciones, Electrónica y Computación). Through these projects, the IAR contributes to the development and optimization of telecommunications and RF systems, ensuring compliance with industry standards and performance requirements.
    
    \item{\textbf{Derived Products and Services (Spin-Offs).}} The fourth working line for technology transfer at IAR involves the creation of derived products and services that result as spin-offs from previous scientific research and technological developments in collaboration with SMEs and government agencies. These derived products and services are the outcome of open innovation initiatives, where IAR's expertise and technologies are applied in novel ways to address social and industry related challenges and opportunities.
\end{itemize}

% xxxxxxxxxxxxxxxxxxxxxxxxxxxxxxxxxxxxxxxxxxxxxxxxxxxxxxxxxxxxxxxxxxxxxxxxxx
\section{Understanding the relationship with SMEs}
\label{sec:TTSME}
It is utterly important to recognize and understand the place that SMEs take as drivers of economic growth and employment. Therefore, one of IAR’s missions is to actively engage with SMEs to identify their needs and challenges, providing tailored solutions and access to its specialized resources and facilities. By understanding the issues faced by SMEs and leveraging its strengths, IAR aims to create meaningful collaborations that promotes technology adoption, supports local industry development, and contributes to the advancement of the Argentine scientific and technological ecosystem.

\subsection{Identifying Needs and Issues in SMEs}
SMEs in Argentina face a range of challenges due to the country's economic and political landscape. The high annual inflation rate, exceeding 70 \% in the last five years, has put considerable strain on businesses' financial stability. Moreover, restrictions on imports and foreign currency, coupled with higher taxes on imported goods, have complicated the procurement process and access to even basic technologies.

The lack of credit for production hampers SMEs' ability to invest in R\&D activities. To address these issues, SMEs often seek solutions that focus on imports substitution, enabling them to develop custom products and services that cater to the local market. Access to specialized facilities, equipment, and technical expertise is a game changer in enhancing their competitiveness and product offerings.

\subsection{IAR's Weaknesses and Strengths}
One of the most significant challenges organizations face is the attraction and retention of a skilled workforce. In the scientific and technological fields, the competition for researchers and professionals can be fierce, and IAR, like any other research institution, is influenced by the political and economic climate when it comes to recruiting and retaining talent. Presently, numerous young researchers and engineers are drawn abroad or to the private sector due to the highly competitive salary offers and favorable conditions they receive, making it difficult for research institutions to retain top talent.

While IAR offers valuable services, competitive pricing with high running costs can be a concern for SMEs with limited budgets. Additionally, the procurement and administrative processes are usually slower compared to the private sector, which can impede prompt responses to market demands.

Furthermore, limited incentives for technology transfer can pose challenges in engaging SMEs, as they may be cautious about investing in new technologies without clear incentives or cost-sharing arrangements.

However, IAR boasts a collaborative, flexible, and dynamic team, committed to finding creative solutions to meet the specific needs of SMEs. The support from graduate engineering students, who contribute fresh ideas and energy to projects, further strengthens its capabilities.

Geographically, IAR's strategic location offers a significant advantage, as it enables close collaborations with regional industries and research units at the heart of the country's economic development. The Institute's wide network of research units and technical support enhances its capabilities by providing access to diverse expertise and resources. Notably, these collaborations often involve renowned organizations such as the National University of La Plata (UNLP) at the local level, the Scientific Research Commission of the Buenos Aires Province (CIC) at the regional level, and CONICET, which boasts a nationwide network of research units.

% xxxxxxxxxxxxxxxxxxxxxxxxxxxxxxxxxxxxxxxxxxxxxxxxxxxxxxxxxxxxxxxxxxxxxxxxxx
\section{Future Plans and Collaboration}
\label{sec:TTfuture}
In pursuit of its mission to drive technology transfer with significant social and productive impact, IAR is actively seeking strengthened collaborations and expanded partnerships at both local and international levels. Furthermore, the Institute is committed to enhancing cooperation with various organizations, research institutions, and industries to create a more robust and virtuous ecosystem.

One of the key strategies for the future is to establish and reinforce Public-Private Partnerships (PPPs) with commerce and productive chambers, facilitating the exchange of knowledge and resources. Recent business trips, including the initiative from the Polish Embassy, collaborations with ALCE (Agencia Latinoamericana y Caribeña del Espacio), and partnerships with esteemed organizations like SAO (Smithsonian Astrophysics Observatory) for the ngEHT digital back-end, AFORS (Air Force Office of Scientific Research), and ONRG (Office of Naval Research Global), offer valuable opportunities to explore new avenues of collaboration and access exceptional expertise.

As part of its commitment to supporting tech-based companies, IAR will continue to actively engage with SMEs, providing tailored solutions and technological services to address their unique challenges. By granting access to specialized facilities, equipment, and a skilled workforce, the Institute aims to empower tech-based companies and contribute to their growth and success. To achieve these ambitious goals, IAR seeks to expand its workforce while maintaining and enhancing existing capabilities.

It is essential as well to address gender parity within the organization. Currently, the female technical staff at IAR is underrepresented, emphasizing the need for measures to promote diversity and inclusion in the workforce. By adopting an inclusive environment and implementing initiatives to attract and retain female talent in technical roles, it is possible to reach a broader pool of expertise and perspectives.

Looking ahead, IAR aims to explore new horizons and reach into untapped sectors with its technology transfer initiatives. For instance, in the medical domain, the ITM project, along with the emerging collaboration with the Hospital de Niños de La Plata to automate the production of baby's feeding bottles, holds the promise of potentially benefiting healthcare practices and patient outcomes.

Additionally, IAR is committed to making significant contributions to the rapidly evolving NewSpace sector, with initiatives oriented towards developing highly demanded communication modules and SAR antennas.

The industrial sector is also a targeted, by expanding the development of IIoT solutions. With a focus on process-oriented approaches and continuous improvement, the Institute is planning to establish new labs encompassing IoT technologies and the industry 4.0 paradigm to support local industries.

This strategic vision adopts a Model-Based Systems Engineering (MBSE) perspective, aiming to optimize technology transfer processes and enhance the quality of its services. By adopting best practices and proven methodologies, the Institute will be able to deliver efficient and effective solutions to its partners, fostering a thriving ecosystem of innovation and collaboration.

% xxxxxxxxxxxxxxxxxxxxxxxxxxxxxxxxxxxxxxxxxxxxxxxxxxxxxxxxxxxxxxxxxxxxxxxxxx
\section{Conclusion}
\label{sec:conclusion}
The Instituto Argentino de Radioastronomía stands as a prominent institution in Argentina, driven by a steadfast commitment to scientific exploration and technological advancements. Over its rich history spanning more than 60 years, the IAR has made significant contributions to the field of radio astronomy, shaping the country's position in the South American and international scientific community.

In parallel to its contributions to radio astronomy research, the IAR has also actively pursued technology transfer initiatives to extend the impact of its developments beyond its scope. Leveraging the multidisciplinary nature of radio astronomical research, the IAR has identified and oriented transformative technologies with versatile applications across various industries and domains.

Throughout its 25-year journey in technology transfer, the IAR has continuously evolved, adaptively responding to the dynamic political, economic, and technological landscape. From its early beginnings without a formally constituted Technology Transfer Office or Area, the Institute provided crucial services in response to the aftermath of the 2001 crisis. This marked a significant step in laying the groundwork for formal technology transfer activities. Between 2005 and 2015, the IAR solidified its commitment to technology transfer, undertaking complex and demanding projects that contributed to strategic advancements. Since 2018, the Institute has entered an expansion phase, forging new collaborations and partnerships, and amplifying its impact on society. The IAR has demonstrated resilience and dedication, driving positive change with meaningful contributions to science, technology, and society at large.

Central to its efforts lays the active collaboration with small and medium-sized enterprises (SMEs), recognizing their role in driving economic growth and employment. By understanding the unique needs and challenges faced by SMEs, the IAR has provided tailored solutions, access to specialized resources, and facilities, supporting technological adoption and local industry development.

For the coming years, IAR foresees a strengthened collaboration and expanded partnerships, both at the local and international levels. Public-Private Partnerships (PPPs) with commerce and productive chambers will be a key strategy, facilitating knowledge exchange and resource-sharing amidst the constraints of the current landscape. The Institute's unwavering support for tech-based companies will empower them to thrive, while the pursuit of untapped sectors such as medical, NewSpace, and industrial domains will open new frontiers for technology transfer.

To ensure continued excellence in its technology transfer endeavors, the IAR seeks to expand its workforce, enhance capabilities, and adopt proven methodologies, optimizing processes and delivering efficient, effective solutions.

In conclusion, the Instituto Argentino de Radioastronomía's dedication to technological advancement and collaboration serves as a beacon of scientific and technological progress in Argentina. As it continues to navigate the challenges and opportunities of the future, the IAR's legacy of knowledge-intensive developments and commitment to making a positive societal impact will continue to thrive in the world of radio astronomy and beyond.

% xxxxxxxxxxxxxxxxxxxxxxxxxxxxxxxxxxxxxxxxxxxxxxxxxxxxxxxxxxxxxxxxxxxxxxxxxx
\vspace{0.3cm}

{\bf Acknowledgements:} The authors would like to extend their sincere gratitude to the following individuals and groups who have contributed to the development and review of this article:
The Directorate of the IAR, Dr. Gustavo E. Romero, Director of IAR, and Dra. Paula Benaglia for their support and contributions to the technology transfer endeavors of the institute.
The IAR's Technology Sector, comprising the following members, for their dedication and active participation in technology transfer activities: Guillermo Gancio, Augusto Donantueno, Carlos Cristina Miguel, Emiliano Rasztocky, Pablo Alarcón, Santiago Spagnolo, Daniel Perilli, Julián Galván, Hugo Command, Evelina Tarcetti, Ruben Morán, Eliseo Díaz, Facundo Aquino, Matías Contreras, Luis Ferrufino, Gastón Valdéz, Juan Manuel González, Darío Capucchio, Marcos Borgetto, Lucía Benaglia, Lucía Bagnato and Claudia Boeris.
Researchers Dr. Manuel Fernández and César Caiafa for their valuable contributions to the development of technology transfer initiatives at IAR.
To our pioneers, Eng. Juanjo Larrarte and Juan Sánz for their effort, perseverance and commitment, which allowed the IAR to be located on the map.
Special thanks to Prof. M.~Sc.~Econ. Ana Luz Abramovich for reviewing this article and providing constructive feedback to enhance its quality and clarity.

\end{document}